\documentclass[journal,twoside,web]{ieeecolor}
\usepackage{generic}
\usepackage{bm}
\usepackage{amsmath,amssymb,amsfonts}
\usepackage{algorithmic}
\usepackage{graphicx}
\usepackage{algorithm,algorithmic}
\usepackage{hyperref}
\hypersetup{hidelinks}
\usepackage{textcomp}
\usepackage{threeparttable}
\usepackage[backend=biber, style=ieee, maxcitenames=2, mincitenames=1]{biblatex}
\usepackage{bbm}
\usepackage{placeins}
\newcommand{\citet}[1]{\Citeauthor{#1} \cite{#1}}
\def\BibTeX{{\rm B\kern-.05em{\sc i\kern-.025em b}\kern-.08em
    T\kern-.1667em\lower.7ex\hbox{E}\kern-.125emX}}
\begin{document}
\title{ManiNeg: Manifestation-guided Multimodal Pretraining for Mammography Classification}
\author{Xujun Li, Xin Wei, Jing Jiang, Danxiang Chen, Wei Zhang, and Jinpeng Li, \IEEEmembership{Member, IEEE}
\thanks{Li, X., Jiang, J., Chen, D, and Zhang, W. are with Department of Oncology, Ningbo NO.2 Hospital. They are also with Department of Breast Surgery, Ningbo NO.2 Hospital, Ningbo 315000, China.}
\thanks{Wei, X. and Li, J. are with Ningbo Institute of Life and Health Industry, University of Chinese Academy of Sciences, Ningbo 315010, China.}
\thanks{This work was supported by National Natural Science Foundation of China (62106248), Project of Ningbo Leading Medical and Health Discipline (2022-B13), Ningbo Clinical Research Center for Medical Imaging (2021L003), and Provincial and Municipal Co-construction Key Discipline for Medical Imaging (2022-S02).}
\thanks{Corresponding author: Jinpeng Li (lijinpeng@ucas.ac.cn).}}

\maketitle

% 检查统一用语：Mammography，作形容词时使用形容词形式mammographic，表示乳腺钼靶图片时使用mammogram。只有在单独表示乳腺钼靶摄影术时使用Mammography
% 240105update:表示乳腺钼靶图片时有两种形式，分别是mammogram和mammographic image. 两者视情况可能不需要完全统一，但倾向于使用mammogram
% batch size，有空格
% minibatch，有空格无连字符
% pretrain，无空格无连字符
% multimodal不带连字符
% 关于数据，区分case, instance等词汇
% 可能的补充实验：1.大batch size+标准simclr 2.有监督预训练，01label
% 检查：模板中要求公式不带Eq.的前缀，除非在句首。最后可统一检查这一点。

\begin{abstract}
Breast cancer is a significant threat to human health. Contrastive learning has emerged as an effective method to extract critical lesion features from mammograms, thereby offering a potent tool for breast cancer screening and analysis. A crucial aspect of contrastive learning involves negative sampling, where the selection of appropriate hard negative samples is essential for driving representations to retain detailed information about lesions. In contrastive learning, it is often assumed that features can sufficiently capture semantic content, and that each minibatch inherently includes ideal hard negative samples. However, the characteristics of breast lumps challenge these assumptions. In response, we introduce ManiNeg, a novel approach that leverages manifestations as proxies to mine hard negative samples. Manifestations, which refer to the observable symptoms or signs of a disease, provide a knowledge-driven and robust basis for choosing hard negative samples. This approach benefits from its invariance to model optimization, facilitating efficient sampling. To support ManiNeg and future research endeavors, we developed the MVKL dataset, which includes multi-view mammograms, corresponding reports, meticulously annotated manifestations, and pathologically confirmed benign-malignant outcomes. We evaluate ManiNeg on the benign and malignant classification task. Our results demonstrate that ManiNeg not only improves representation in both unimodal and multimodal contexts but also shows generalization across datasets. The MVKL dataset and our codes are publicly available at https://github.com/wxwxwwxxx/ManiNeg.

\end{abstract}


% 注意统一用语：cross enrtopy,没有连字符
% 注意统一用语：最好最后检查所有连字符

\section{Introduction}
\label{sec:introduction}
\IEEEPARstart{B}{reast} cancer remains a formidable challenge to global health. For example, it constitutes one-third of all new cancer cases among women in the US, making it the second most lethal cancer among women, trailing only behind lung cancer \cite{siegel2023cancer}. Mammography, as an early screening method, has proven effective in lowering the mortality rate associated with breast cancer. Extracting critical information from mammograms is crucial for enhancing breast cancer detection rates and further analyzing breast lumps. Contrastive learning, a powerful deep learning-based method for extracting representations, has gained prominence. Originating from unsupervised learning, it discerns whether two image views belong to the same instance, offering an advantage over supervised learning by obviating the need for labels and yielding more robust, generalizable representations due to its non-task-specific training approach \cite{wang2020understanding}. Its advantageous properties have led to its expansion into multimodal and supervised variants.

\begin{figure}[t]
    \centering
    \includegraphics[width=1.0\linewidth]{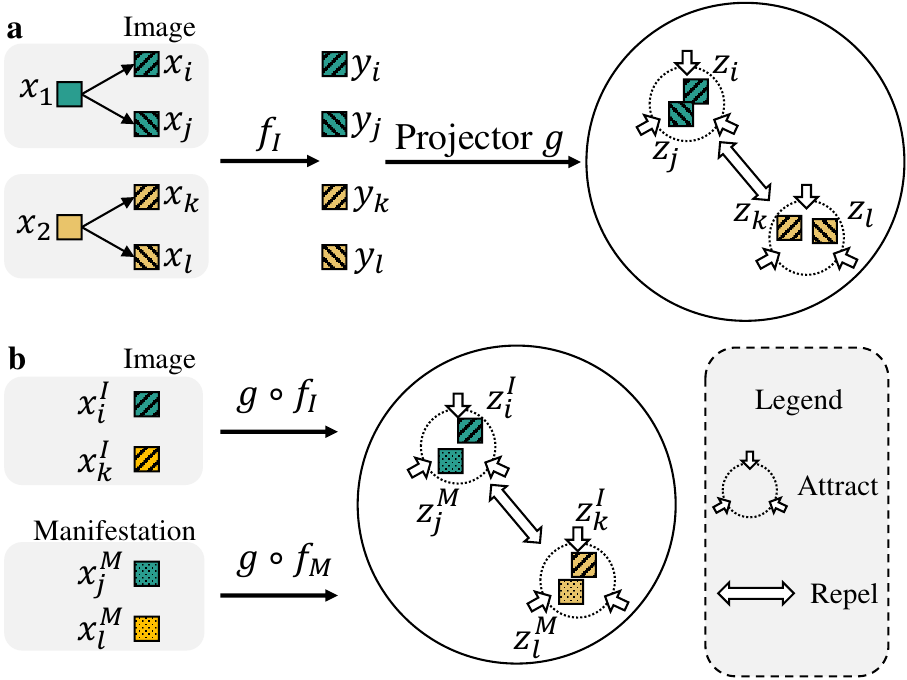}
    \caption{Contrastive pretraining scheme for mammography analysis. (a) Unimodal learning with images. (b) Multimodal learning with images and manifestations. $i$ and $j$ denote a pair of views from instance $\bm x_1$, marked in green. $k$ and $l$ denote a pair of views from instance $\bm x_2$, marked in yellow. $f_I(\cdot)$ and $f_M(\cdot)$ represent the encoders for images and manifestations, respectively. $g(\cdot)$ is the shared projector. Samples are drawn closer to the positive samples (e.g., $\bm z_i$ and $\bm z_j$ attract each other) and are repelled from negative samples (e.g., $\bm z_i$ and $\bm z_k$ repel each other) in the representation sphere. Based on these, the model learns features from the instances in an unsupervised manner.}
    \label{fig:model}
\end{figure}

 As Fig.~\ref{fig:model} shows, contrastive learning involves creating pairs of views from an instance through data augmentation or cross-modal matching, then sampling a set of these views into minibatches. Then, views from the same instance are designated as positive samples, while those from different instances are negative samples. The model minimizes the distance between positive sample representations and maximizes the distance between negative ones, utilizing a loss function to highlight essential features of the instance.

Negative sampling is pivotal in contrastive learning, influencing the differentiation between positive and negative samples. Hard negative samples, i.e., semantically similar but distinct from positive samples, encourage the model to explore semantic differences, leading to the extraction of informative and aligned representations. SimCLR ~\cite{chen2020simple} employs the uniform sampling to select hard negative samples with a broad array of candidates. This method, laying the groundwork for many studies, relies on two key assumptions: First, the representations closely correspond to the semantic content of interest, allowing their use in hard negative sample selection. Second, each batch likely contains hard negative samples, which the cross-entropy calculation can then emphasize. In large datasets with a focus on image saliency and common practice of large batch sizes, these assumptions are generally applicable.

However, the small size and obscured nature of breast lumps, challenge these assumptions. Unlike in scenarios where image saliency is assumed, the alignment between representations and breast lump characteristics during training is not guaranteed, and the representations themselves evolve as the model is optimized. This complicates hard negative sample selection, traditionally reliant on random batch sampling or memory banks \cite{chen2020simple, he2020momentum}, as sampling across the entire dataset at each training step is computationally prohibitive. Moreover, the skewed distribution of mammographic data and constraints on batchsize due to dataset scale limit the efficacy of contrastive learning approaches like SimCLR. With the evolution of contrastive learning towards multimodal domains, additional modalities present a promising avenue to circumvent these constraints. In the image-text contrastive learning, text, when accurately describing corresponding images, aligns closely with the targeted semantics and remains constant throughout the training process, thereby indicating its potential as a proxy for semantic-guided hard negative sampling.

% 关于训练效率的问题：在这里的文段附近，可以考虑添加关于训练效率的问题，即基于manifestation，我们**只**选择了难负样本，这会很大地优化训练效率
% 240105：一个额外的点：可以明确写出，manifestation表达的是我们interest或concern的特征

The selection of hard negative samples necessitates a well-defined metric to quantify semantic discrepancies between instances. While text similarity comparison offers a plausible approach, its complexity and the nuanced nature of textual similarity render it less effective for the stringent requirements of hard negative sampling analysis. On the contrary, in medical imaging analysis, manifestations emerge as a superior proxy. Manifestations, encapsulating the symptoms and signs of a disease in a structured format, are pivotal for diagnosis, maintaining a direct and significant correlation with the lesions of interest. The structured nature of manifestations simplifies the semantic distance measurement, enabling the evaluation of semantic similarity between instances through a straightforward application of the Hamming distance. 

% 关于这一段内容：后续可以补充实验：manineg与大batchsize的对比。如果这样的实验里manineg有优势，则可以在下一段添加
% “我们可以在较小的batchsize下达到（超越）大batchsize的效果”

Building on these insights, we introduce \textbf{ManiNeg}, a novel manifestation-guided hard negative sampling strategy tailored to effectively navigate the challenges posed by representation-based approaches and the characteristics of mammography images. Empirical studies validate that ManiNeg significantly enhances representation learning in both unimodal and multimodal settings, and the generalization across datasets.

We have developed the \textbf{M}ammography \textbf{V}isual-\textbf{K}nowledge-\textbf{L}inguistic (MVKL) dataset. This dataset mirrors the authentic data landscape encountered in medical practice. Each case in the MVKL dataset encompasses three modalities: mammography images, detailed manifestations, and radiology reports, all accompanied by pathologically confirmed benign-malignant outcomes and pixel-level annotations of breast lumps. Through the creation and annotation of extensive manifestation tables, we facilitate direct evaluation of ManiNeg's impact. We make the MVKL dataset publicly accessible, aiming to foster continued research in this and related areas.

% 检查240103：讲到mammographic images的地方可以添加multi-view
% 检查240105：注意统一用语，乳腺钼靶图片有两种说法，分别是mammographic images和mammogram。倾向于使用mammogram
% 检查240105：注意统一用语，关于报告：Intro中有提到病历（medical report）和影像报告（imaging report）两种医疗文书。后续检查需要确定两件事
% 1. 统一用语，所有的病历和医疗报道都应该使用这两种用语。特别是后文介绍数据库时需要强调imaging report而不是只使用report
% 2. 确定用语是否规范，后续可以修改这两个用语。例如，倾向于使用Radiology Report代替Imaging Report
% 检查240104：采样分布（抽样分布）可能有其他含义的特指，待检查

To summarize, our primary contributions are manifold:
\begin{itemize}
    \item[$\bullet$] We critically evaluate the conventional hard negative sampling methods prevalent in contrastive learning, highlighting their inadequacies for mammographic data analysis, and advocate for the use of manifestations as a viable proxy to surmount these challenges.
    \item[$\bullet$] We introduce the ManiNeg framework, designed to optimize hard negative sampling through a strategic distribution and methodological approach, grounded in lesion manifestations. Our rigorous testing underscores ManiNeg's practicality and efficacy.
    \item[$\bullet$] We have meticulously compiled and will disseminate the MVKL dataset, featuring multi-view mammograms, corresponding radiology reports, pathologically validated benign-malignant labels, and unique manifestations, to support a wide range of innovations for the community.
\end{itemize}

Section 2 reviews literature on deep learning for mammography, contrastive learning, and negative sampling. Section 3 introduces the MVKL dataset, including its organization, annotation, modalities, and manifestation tables. Section 4 details the ManiNeg method, covering its design and implementation during training. Section 5 validates ManiNeg's effectiveness through various tests, assessing its impact on contrastive learning. Section 6 discusses ManiNeg's limitations and future prospects. The article concludes with Section 7, summarizing our findings and contributions.

\begin{figure*}[ht]
    \centering
    \includegraphics[width=1.0\linewidth]{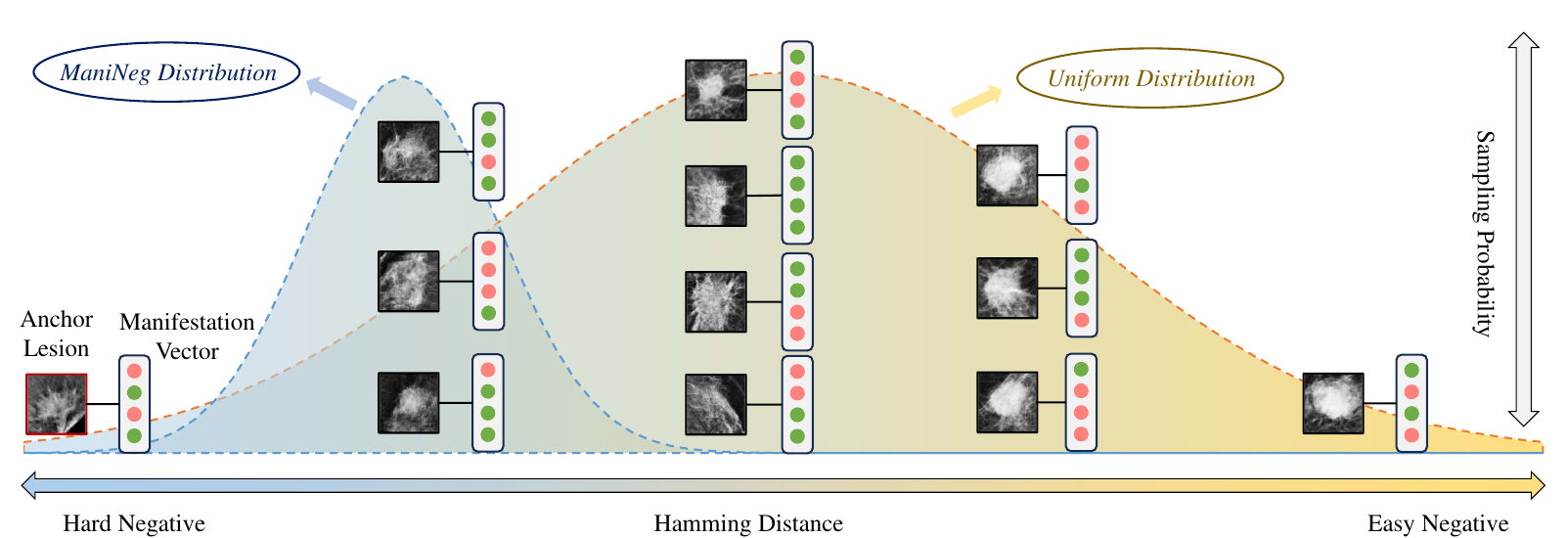}
    \caption{ManiNeg and uniform sampling. We assume independent traits in the manifestation have a 50\% occurrence probability for illustration. On a 4-bit manifestation scenario, where the distribution of Hamming distances between negatives and the anchor aligns with the binomial distribution $B(4,0.5)$. Uniform sampling most frequently results in a Hamming distance of 2, missing the ideal hard negatives at a distance of 1. ManiNeg, sampling by Hamming distance, directly targets these hard negatives, effectively enhancing sample selection for improved learning.}
    \label{fig:main}
\end{figure*}
    
\section{Related Works}
\label{sec:related_works}
% Mammography Image Analysis. While pathology slides are the gold standard for diagnosing breast cancer, with several deep learning studies focused on their analysis [40, 41], mammography remains an essential tool in breast cancer screening. A variety of research has developed multi-scale, multi-view models for detecting breast masses in mammography [42, 43]. For instance, AGN employed a bipartite graph convolutional network alongside an inception graph convolutional network, enhancing correspondence reasoning by integrating information from ipsilateral and bilateral views [43]. Also, in light of the limited availability of labeled data, weakly-supervised learning approaches such as multi-instance learning have shown promise in improving sensitivity, utilizing labels of varying granularity [44].

% A:Efficient breast cancer mammograms diagnosis using three deep neural networks and term variance
% B:Applying Deep Learning Methods for Mammography Analysis and Breast Cancer Detection
% C:Adapting the pre-trained convolutional neural networks to improve the anomaly detection and classification in mammographic images
\textbf{Mammography Image Analysis.} 
Mammography is the foremost screening tool for breast cancer due to its efficiency and effectiveness. Yet, the precision in analyzing breast lumps, particularly those that are concealed, remains a formidable challenge in radiology. This challenge has catalyzed a surge in research exploring the enhancement of mammographic examination through advanced deep learning techniques. \citet{elkorany2023efficient} leverages multiple models to distill features from mammograms, employing a Term Variance feature selection algorithm to isolate the most pertinent features. \citet{prodan2023applying} introduces a novel data augmentation strategy utilizing StyleGAN \cite{karras2019style} to generate additional data. \citet{saber2023adapting} adapts models pre-trained on ImageNet to a Long Short-Term Memory (LSTM) network, facilitating the extraction of features. Given that mammograms encompass multiple perspectives, the integration of multi-scale and multi-view features has been proven effective for a more nuanced analysis of breast lumps \cite{cao2019deeplima}. Expanding on this, \citet{liu2021act} employs an array of graph convolutional networks to amalgamate features from both unilateral and bilateral mammographic views. Moreover, semi-supervised techniques, such as multiple instance learning, have demonstrated potential in mammogram analysis under scenarios with limited labels \cite{kallenberg2016unsupervised}.

\textbf{Contrastive Learning.} 
Contrastive learning, a cornerstone in the realm of self-supervised learning, has progressively found application in supervised and multimodal learning contexts \cite{khosla2020supervised}. SimCLR \cite{chen2020simple}, a foundational contrastive learning framework, meticulously examines the influence of data augmentation, network architecture, and loss functions on model efficacy. It notably underscores the critical role of the negative sample size (i.e., batch size) on model performance. MoCo \cite{he2020momentum} introduces a momentum encoder and a memory bank mechanism to enlarge the negative sample pool. SwAV \cite{caron2020unsupervised}, diverging from direct optimization among samples, clusters them to prototypes to optimize sample-to-prototype distances.

\textbf{Multimodal Contrastive Learning.} The versatility of contrastive learning, unconstrained by modality homogeneity, paves the way for its extension into multimodal domains. CLIP \cite{radford2021learning} epitomizes this transition by utilizing separate encoders for images and texts, aligning their representations via a contrastive loss. Various scales of ResNet \cite{resnet} and ViT \cite{dosovitskiy2020vit} function as image encoders, while the Transformer architecture \cite{vaswani2017attention} serves as the text encoder. VSE++ \cite{faghri2018vse++} enhances model performance through a hard negative mining strategy. ViLT \cite{kim2021vilt} merges patch-based image embeddings with text embeddings via a transformer. BEiT3 \cite{beit3} demonstrates the efficacy of discrete image tokens, generated through VQ-VAE \cite{van2017neural}, in improving performance. ALBEF \cite{li2021align} introduces a pre-modal fusion alignment loss, fostering improved intermodal relationship learning. VLMo \cite{bao2021vlmo} incorporates modality-specific modules within the Transformer, alternating between self-attention and modality expert modules to integrate and learn features across modalities. BLIP \cite{li2022blip} trains a captioner to recognize data noise, allowing for pretraining on large-scale, noisy datasets. PaLI \cite{chen2022pali} extends unimodal text and image models for text generation tasks, culminating in scalable multimodal models. REFERS \cite{zhou2022generalized} aligns features extracted from X-Ray images with corresponding radiology reports through contrastive learning, effectively leveraging report data. \citet{hager2023best} further validates the utility of tabular data in enhancing multimodal contrastive learning approaches.

% A: Contrastive Learning with Hard Negative Samples
% B: On Negative Sampling for Contrastive Audio-Text Retrieval
% C: CONDITIONAL NEGATIVE SAMPLING FOR CONTRASTIVE LEARNING OF VISUAL REPRESENTATION
% D: Exploring the Impact of Negative Samples of Contrastive Learning: A Case Study of Sentence Embedding
% E: Hard Negative Mixing for Contrastive Learning
% F：CONTRASTIVE LEARNING WITH HARD NEGATIVE SAMPLES
% G: Hard Negative Sampling Strategies for Contrastive Representation Learning
% H：Filtering, Distillation, and Hard Negatives for Vision-Language Pre-Training
% I：Robust Contrastive Learning Using Negative Samples with Diminished Semantics
% J: Semantically-Conditioned Negative Samples for Efficient Contrastive Learning

\textbf{Hard Negative Sampling.}
The concept of contrastive loss incorporates a weighting mechanism for hard negative samples via cross-entropy, thereby legitimizing uniform random sampling as an effective strategy for hard negative sample selection \cite{chen2020simple}. Progressing from this foundation, numerous studies have delved into optimizing the selection of hard negative samples. MoCo \cite{he2020momentum} broadens the negative sample spectrum with its memory bank mechanism. Concurrently, \citet{cao2022exploring} delve into refining memory bank size optimization through a maximum traceable distance metric. \citet{wu2020conditional} innovatively selects hard negative samples by gauging mutual information between representations and sampling within a specified proximity to the positive anchor. Drawing inspiration from data mixing techniques, \citet{kalantidis2020hard} introduces a representation-driven hard negative mixing strategy to dynamically fabricate hard negative samples. \citet{robinson2020hard} conceptualizes a latent class distribution alongside a von Mises-Fisher distribution, centered around positive samples, to differentially weight hard negative samples across distinct classes via importance sampling. DiHT \cite{radenovic2023filtering} further extrapolates this approach to the multimodal sphere. \citet{tabassum2022hard} amalgamates anchor similarity, model uncertainty, and representativeness in the hard negative sample selection calculus to enhance sample quality. \citet{ge2021robust} innovates non-semantic negative sample generation through patch-based and texture-centric augmentations, steering model focus towards semantic discernment. \citet{neill2021semantically} predicates negative sampling on semantics, inferred through pretrained representations. VSE++ \cite{faghri2018vse++} employs a triplet-based loss function, prioritizing the learning influence of the most challenging negative samples, a technique also embraced by ALBEF \cite{li2021align} and VLMo \cite{bao2021vlmo}.

In summary, most existing studies presume two fundamental assumptions about negative sampling: a strict alignment between representations and semantics, and the inherent presence of hard negative samples within each mini-batch. Such presuppositions suggest that traditional methods might falter in specialized contexts like mammography analysis, where these assumptions are less applicable. ManiNeg introduces an innovative solution by leveraging an auxiliary modality as a semantic proxy, directly addressing these challenges.

\begin{figure}[t]
    \centering
    \includegraphics[width=1.00\linewidth]{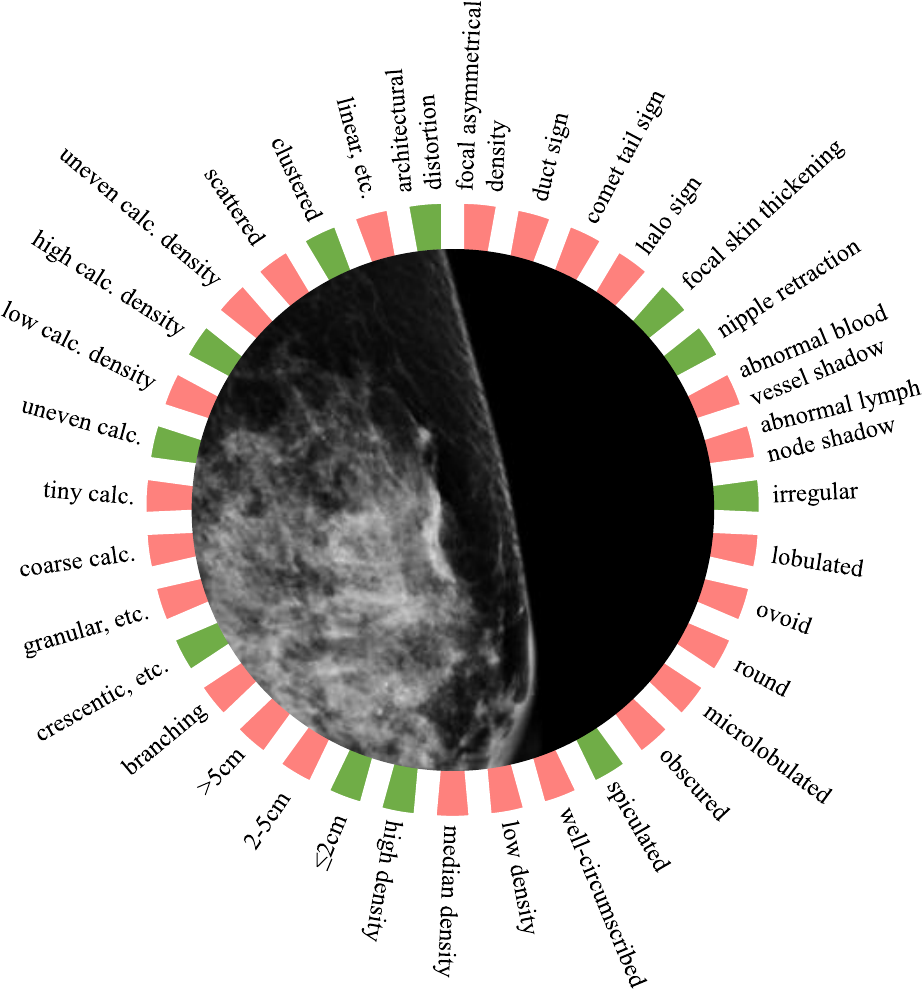} 
    \caption{Schematic of a binary manifestation vector. The manifestation is initially annotated according to the header listed in Table~\ref{tab:mani_header} and subsequently expanded into a 35-dimensional binary vector. We show a binary manifestation vector surrounding its corresponding breast lump. Green: presence; Red: absence.}
\label{fig:mani_demo}
\end{figure}

\section{Dataset}
\label{sec:dataset}
% - 乳腺钼靶介绍
% - Manifestation介绍
% - 数据集概况
% - 整理过程
% - 隐私声明
We have curated and meticulously annotated the first multimodal mammographic dataset, termed the \textbf{M}ammography \textbf{V}isual-\textbf{K}nowledge-\textbf{L}inguistic dataset (MVKL), to provide a holistic view of mammographic analysis. 

\textbf{Mammography.} 
% The high incidence of breast cancer makes early diagnosis and treatment an urgent necessity, and mammography serves as such an effective screening and diagnostic tool for breast diseases.
Mammography captures the breast's X-ray images to identify and analyze suspicious high-density areas indicative of breast lumps. To ensure a thorough examination of these lumps, a standard mammogram for a unilateral breast typically encompasses two conventional views: the \emph{mediolateral oblique} (MLO) and the \emph{craniocaudal} (CC) views. These diverse physical perspectives are crucial for contrastive learning, as they enable the model to discern essential features by comparing different views. Accordingly, we preserved all original views from each examination to enrich the dataset's utility for learning algorithms.

\textbf{Manifestations.}
Manifestations, indicative of disease symptoms and signs, serve as a fundamental diagnostic cornerstone. They originate from patient complaints and clinical findings, characterized by their structured nature. Clinicians and radiologists often document critical manifestations in medical reports, facilitating the extraction of such data through Natural Language Processing (NLP) techniques. For instance, the Chest14 dataset \cite{wang2017chestx} utilized NLP to gather various diagnostic labels from chest X-ray data. Nonetheless, NLP methods may introduce noise, and potentially insightful manifestations occasionally omitted by radiologists could provide valuable information. To address these challenges, we developed a generic table format to systematically catalog breast lump characteristics, ensuring comprehensive case annotation. 
% This eliminates worry about the noise in the manifestations, allowing us to focus on the improvements introduced by them. 
% It is worth noting that using the aforementioned noisy manifestations is not entirely unfeasible, but it requires additional steps to assess the impact of the potential noise. We will discuss this topic in detail in the discussion section and leave it as a related work.

The manifestations encompasse 17 principal traits categorized into four mass traits, four calcification traits, and nine miscellaneous traits, as detailed in Table~\ref{tab:mani_header}. The manifestations are transformed into a 35-dimensional binary vector, each dimension reflecting the presence or absence of specific traits. Fig.~\ref{fig:mani_demo} shows a schematic diagram of such a manifestation binary vector. This binary vector not only simplifies model processing but also enables the assessment of case similarities through the Hamming distance, offering a straightforward and effective means for comparing mammographic cases.

\renewcommand{\arraystretch}{1.0}
\begin{table}[htp]
    \caption{The manifestations. The miscellaneous options are independent, and the rest options are mutually exclusive.}
    \centering
    \begin{tabular}{p{2.4cm}|p{5.6cm}}
    \hline\hline
        \textbf{Manifestations} & \textbf{Options} \\ \hline
        \textbf{mass shape} & irregular, lobulated, ovoid, round  \\ \hline
        \textbf{mass edge} & microlobulated, obscured, spiculated,\newline well-circumscribed \\ \hline
        \textbf{mass density} & low, median, high  \\ \hline
        \textbf{mass size} & $\leq$2cm, 2-5cm, $>$5cm  \\ \hline
        \textbf{calcification shape} & branching, crescentic/annular/gritty/thread-like,\newline granular/popcorn-like/large rod-like/eggshell-like  \\ \hline
        \textbf{calcification size} & coarse, tiny, uneven  \\ \hline
        \textbf{calcification density} & low, high, uneven  \\ \hline
        \textbf{calcification\newline distribution} & scattered, clustered, linear/segmental  \\ \hline
        \textbf{miscellaneous} & architectural distortion, focal asymmetrical\newline density, duct sign, comet tail sign, halo sign, focal skin thickening/retraction, nipple retraction, abnormal blood vessel shadow, abnormal lymph node shadow \\ \hline\hline

    \end{tabular}
    \label{tab:mani_header}
\end{table}
% 240105注意统一用语：BI-RADS始终对应categories
% 注意下一段中的用语，imaging可能要替换为radiology，注意不要遗漏
% 240105注意统一用语：MVKL，不是MKVL或其他顺序

\textbf{Radiology Reports.}
While the textual modality is not the primary focus of our current research, it plays a critical role in multimodal studies, and thus, radiology reports form an integral part of the MVKL dataset. These reports, penned by radiologists, succinctly detail the location of breast lumps, essential manifestations, and the Breast Imaging Reporting and Data System (BI-RADS) categories in a concise, free-text format. Notably, BI-RADS offers a preliminary benign-malignant classification of lumps based on imaging and patient feedback, frequently aligning closely with pathological outcomes. The BI-RADS framework classifies breast lumps into seven risk levels, from 0 (incomplete) to 6 (proven malignancy), with intermediate categories providing a nuanced risk assessment. Ensuring comprehensiveness, each radiology report within the MVKL dataset is accompanied by a BI-RADS categorization.

% 240105注意统一用语：良恶性结果，可以用benignity and malignancy，也可以用benign and malignant，需要检查语境，考虑是否需要统一用语。
% 240105补充↑：统一使用benign and malignant即可，理由如下：
% 在讨论医学或科学主题，特别是指肿瘤的良性或恶性时，建议使用“benign”作为形容词  ---by ChatGPT.
% 关于是否可以用作名词：
% 在医学文献中，虽然“benign”和“malignant”通常作为形容词使用，用以描述肿瘤或其他医学条件的性质，但在某些情况下，它们也可以作为名词使用。这种用法比较特殊，通常出现在讨论肿瘤类型时，尤其是当上下文已经明确指出正在讨论的是肿瘤或类似的医学状况时。
% ---by ChatGPT.
% 240105，一些地方的“经病理（pathology，pathological等）验证”可以考虑换成“经过活检（biopsy）检查，包含活检结果”等。有时这样的写法更加合理。

% 统一用语：database是不对的，要修改为dataset 

\renewcommand{\arraystretch}{1.5}
\begin{table}[htp]
    \centering
    \setlength{\tabcolsep}{6pt}
    \caption{Label distributions. Ben: Benign; Mal: Malignant.}
    \begin{tabular}{l c c c c c c c c}
        \hline\hline
        Partition & \multicolumn{2}{c}{Training} & &\multicolumn{2}{c}{Validation} & &\multicolumn{2}{c}{Test} \\
        \cline{2-3}\cline{5-6}\cline{8-9}
        Label & Ben. & Mal.& & Ben. & Mal.& & Ben. & Mal. \\
        \hline
        MVKL & 621 & 1325 & &102 & 174 & &158 & 384 \\
        CBIS-DDSM & 1494 & 1083 & &189 & 98 & &428 & 276 \\
        \hline\hline
    \end{tabular}
    \label{tab:label_dist}
\end{table}

\textbf{Labels.} 
All breast lumps documented in the MVKL dataset are corroborated with pathologically confirmed benign or malignant outcomes, serving as the objective ground truth for benign-malignant predictions in our research. 

\textbf{Mask of Breast Lumps.}
All breast lumps are annotated with pixel-based masks on their corresponding mammograms, enhancing the dataset's utility for detailed imaging analysis.

\textbf{Collection Procedure.} 
Focusing on breast lumps as the primary unit of analysis, the MVKL dataset encompasses data from recent clinical studies conducted at Ningbo No.2 Hospital. This includes breast lumps that underwent biopsy examinations, along with their corresponding mammograms and radiology reports. A systematic annotation process was employed, involving physicians who annotated manifestations based on a predefined schema and pixel-based masks. This process unfolded in two stages: initially, five attending and chief physicians performed a collaborative preliminary annotation, leveraging biopsy results and radiology reports to ensure accuracy and reduce annotation burden. Subsequently, a senior chief physician reviewed these annotations to resolve any inconsistencies. Ultimately, the dataset compiled encompasses 2764 breast lumps from 2671 mammographic examinations, with a detailed label distributions provided in Table~\ref{tab:label_dist}.

%%%% 下面这个表也是ViKL抄的。视情况修改。
%%%% 可以考虑和manifestation表头直接合并，形成一个标签分布的条形图。

\textbf{Privacy Statement.} 
Upon data collection, all potentially identifying information was anonymized. This includes removing specific details from Dicom headers, such as names of patients and physicians, dates of birth, examination dates, and identifiers like Accession Numbers, Patient IDs, Study IDs, Study Instance UIDs, Series Instance UIDs, and SOP Instance UIDs, ensuring the privacy and confidentiality of patient data.

\section{Method}
\label{sec:method}
In this section, we first review the paradigm of contrastive learning and analyze its negative sampling scheme. Then we elaborate on the concept and implementation of ManiNeg and explain how it addresses the issue of hard negative sampling. Finally, we illustrate how ManiNeg functions within the training process with an intuitive demonstration.

\subsection{A In-depth Look into Contrastive Learning}
\label{sec:method:clRecap}

% 注意统一用语：对比学习由“自监督学习”发展而来。考虑统一为“自监督学习”，前文有些地方写的是“无监督学习”。
% anchor view $\bm x_i$
Contrastive learning, a self-supervised learning approach, has progressively emerged as a pivotal technique for achieving multimodal alignment. This evolution has seen variations in models and training objectives across different methodologies. For the purpose of clarity and universality, we use SimCLR~\cite{chen2020simple} as a foundational model to elucidate the basic principles of contrastive learning.

\textbf{Basic Framework.} Illustrated in Fig.~\ref{fig:model}, the contrastive learning framework posits the generation of a pair of differing views $\{\bm x_i,\bm x_j\}$ from the same instance $\bm x$. Within this framework, for any given anchor view $\bm x_i$, its counterpart $\bm x_j$ originating from the same instance is tagged as a positive sample and is attracted closer in the representational space. Conversely, views emanating from disparate instances $\{\bm x_k|k \neq i,k \neq j\}$ are designated as negative samples and are distanced. This mechanism of view generation predominantly leverages data augmentation techniques in self-supervised learning contexts.

% 注意检查：所有表示向量的字母都应该标记为黑体，带有\bm标记

SimCLR implements a bifurcated model structure to process and represent data. Initially, a backbone network $f(\cdot)$, exemplified by ResNet50, functions to extract feature representations $\bm y_i$ from input views, denoted as $\bm y_i=f(\bm x_i)$. Subsequently, a MLP, $g(\cdot)$, projects these extracted features into a distinct projection space, represented as $\bm z_i=g(\bm y_i)$. The contrastive learning process predominantly operates within this projection space $\bm z_i$ during the pretraining phase. However, for downstream tasks, the model reverts to utilizing the feature $\bm y_i$ for validation, thereby preserving a richer information spectrum from the initial input within $\bm y_i$. For simplicity and clarity, future references to views, representations, or samples within this discussion will pertain to the projection space vector $\bm z$, unless explicitly indicated otherwise.

The core objective of contrastive learning in its pretraining phase is to discern the positive sample pair amongst all pairs within the same batch for each view. This identification challenge is addressed through a loss function known as \emph{the normalized temperature-scaled cross entropy loss} (NT-Xent):

\begin{equation}
\label{eq:clbase}
\ell(\bm z_i, \bm z_j) = -\log \frac{\exp(\mathrm{sim}(\bm z_i, \bm z_j)/\tau)}{\sum_{k=1}^{2N} \mathbbm{1}_{[k \neq i]}\exp(\mathrm{sim}(\bm z_i, \bm z_k)/\tau)}~.
\end{equation}

Here, $\tau$ represents the temperature parameter, $\mathbbm{1}$ the indicator function, and $\mathrm{sim}(\cdot)$ denotes the similarity measure, which in the context of contrastive learning, is often calculated using cosine similarity, i.e., $\mathrm{sim}(\bm u, \bm v) = \frac{\bm u \cdot \bm v}{|\bm u| |\bm v|}$.

\textbf{Hard Negative Sampling Schemes.} The minibatches during the training process of SimCLR are generated via uniform sampling. SimCLR implicitly weights hard negative samples through the NT-Xent loss function. As can be inferred from its name, NT-Xent is composed of two parts:

\begin{equation}
    \label{eq:ce}
    H(p,q) = -p\log(q)~, p=1~,
\end{equation}
\begin{equation}
    \label{eq:q}
    q(\bm z_i, \bm z_j) = \frac{\exp(\mathrm{sim}(\bm z_i, \bm z_j)/\tau)}{\sum_{k=1}^{2N} \mathbbm{1}_{[k \neq i]}\exp(\mathrm{sim}(\bm z_i, \bm z_k)/\tau)}~,
\end{equation}
where $q(\bm z_i, \bm z_j)$ captures the difference in similarity between positive and negative sample pairs. Specifically, if a negative sample $\bm z_k, k \neq j$ exhibits a high similarity to $\bm z_i$, it significantly contributes to the denominator of $q(\bm z_i, \bm z_j)$, thus playing a crucial role in~\eqref{eq:ce}. Such negative samples, which are challenging to differentiate from the anchor samples, are termed hard negative samples.

% In the section \ref{sec:introduction}, we detailed the implicit assumptions of the default sampling method and the issues it presents when applied to mammographic analysis. Here, we provide a concise summary. The default hard negative sampling scheme based on the similarity between representations, leading to two issues. First, the lack of saliency of breast lumps causes misalignment between representations and semantics, which leads NT-Xent to incorrectly weigh hard negative samples, exacerbating the misalignment in a vicious cycle. The second is the sampling efficiency issue. Limited by the continuous change in representations during training, we cannot find hard negative samples without inferring the model. This means that 
Hard negative sampling encompasses two phases: initially, a minibatch is created via uniform sampling from the dataset. Following the inference of this minibatch, the hard negative samples contained within it are then weighted using NT-Xent. The process of sampling hard negative samples thus displays relatively low efficiency. Consequently, it is inferred that acquiring an adequate number of hard negative samples necessitates a large batch size during training, a finding that aligns with the primary conclusions drawn from SimCLR.

Exploring into sampling efficiency prompts the question: what proportion of a minibatch comprises hard negative samples? Introducing a model-independent proxy to represent the features of interest in the input data can provide insight. Proxies stands for the original entities. In the context of breast lumps, manifestations fulfill the role of such a proxy, enabling the assessment of semantic similarity through the comparison of proxies. Breast lumps with akin manifestations, while similar, also differ semantically, rendering them prime candidates for hard negative samples.
% 注意：良恶性的“性”可以翻译为nature

The MVKL dataset employs binary vectors to symbolize manifestations, allowing the semantic differences between manifestations to be quantified via the Hamming distance. This metric also indicates the challenge level of the samples. Assuming the independence and equal probability of each manifestation dimension, the Hamming distance between an anchor sample and negative samples adheres to a binomial distribution, as depicted by the \emph{uniform} curve in Fig.~\ref{fig:main}.

% 注意统一用语 manifestation的长度最好对应size而不是dimension，dimension可能会混淆

\begin{figure}[ht]
    \centering
    \includegraphics[width=.48\textwidth]{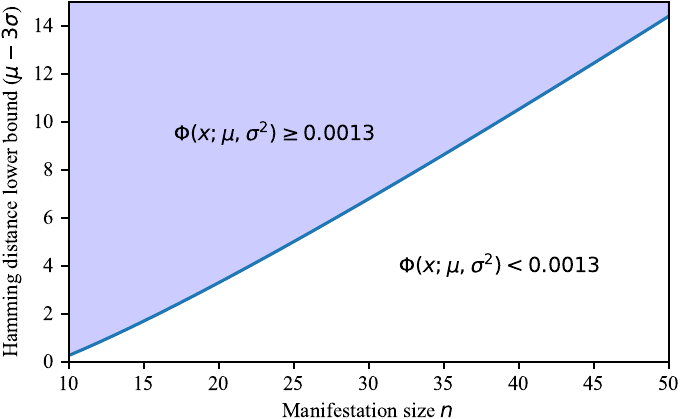}
    \caption{ManiNeg concept in probability. Since each trait within a manifestation is independent with an occurrence probability of $p_m = 0.5$, the Hamming distance $x$ between an anchor and a negative sample's manifestation adheres to a binomial distribution. As the number of traits $n$ increases, this binomial distribution converges towards a Gaussian distribution, $\mathcal{N}(\mu,\sigma^2)$, where its cumulative density function is represented as $\Phi(x;\mu,\sigma^2)$. Based on that, we can delineate the correlation between the manifestation size $n$ and the minimum threshold for sampling Hamming distances, defined as $\mathrm{inf}\{x|\Phi(x;\mu,\sigma^2)\hspace{-0.4em}\geq\hspace{-0.4em}0.0013\}$, effectively $\mu - 3\sigma$. It reveals a direct relationship: as $n$ escalates, the threshold for hard negative sampling—or the lower bound of the Hamming distance—likewise increases. This increment underscores the growing scarcity of hard negative samples with the expansion of manifestation size, highlighting a pivotal challenge in efficiently identifying such samples as the complexity of the data representation increases.}
    \label{fig:lower_bound}
\end{figure}

Hence, when \emph{uniformly} sampling minibatches from the dataset, especially when dealing with multiple semantic features (a common scenario, even in binary classification tasks, where classification often hinges on multiple features), the sampling equates to navigating a \emph{binomial distribution} within the semantic space. According to the Central Limit Theorem, this distribution will approximate a Gaussian distribution characterized by a mean $\mu = np_m$ and a standard deviation $\sigma = \sqrt{np_m(1-p_m)}$, where $n$ represents the manifestation size, and $p_m$ is the probability of a particular trait being positive. Notably, $\mu$ escalates more rapidly relative to $n$ than does $\sigma$, indicating that the procurement of hard negative samples with a close Hamming distance becomes increasingly improbable with more complex data and a greater diversity of features. This concept is visually demonstrated in Fig.~\ref{fig:lower_bound}, suggesting that a uniformly sampled minibatch may contain only a minimal proportion of hard negative samples.

% In conclusion, hard negative sampling scheme that relies on uniform sampling followed by weighting is relatively inefficient. \mathrm{sim}
% 注意统一用语 manifestation中的元素被称作trait，而manifestation作为总称。注意不要把其中的几位称作manifestation

% 在杨辉三角中，当层数足够大时，每一层的频率分布趋近于正态分布。考虑到每一层代表的是二项式系数的分布，我们可以计算出当层数（即试验次数）足够大时，这个分布的均值（mean）和标准差（standard deviation）。

% 1. **均值（Mean）**：对于二项式分布，均值是 \(np\)，其中 \(n\) 是试验次数，\(p\) 是每次试验成功的概率。在杨辉三角的情况下，我们可以将其视为抛硬币的试验，因此 \(p=0.5\)。所以，对于第 \(n\) 层（从0开始计数），均值是 \(n \times 0.5 = \frac{n}{2}\)。

% 2. **标准差（Standard Deviation）**：二项式分布的标准差是 \(\sqrt{np(1-p)}\)。同样地，将 \(p\) 设置为0.5，我们得到标准差 \(\sqrt{\frac{n}{4}}\)。

% 因此，对于杨辉三角的第 \(n\) 层（从0开始计数），其频率分布的均值是 \(\frac{n}{2}\)，标准差是 \(\sqrt{\frac{n}{4}}\)。这意味着随着层数的增加，每一层的分布不仅在形状上越来越像高斯分布，其统计特性（均值和标准差）也遵循简单的数学规则。

\subsection{ManiNeg in Motivation}
\label{sec:method:maniNegIntro}

% In this case, using a larger batch size to find potential hard negative samples as much as possible can be considered the only choice.
Addressing the limitations inherent in the traditional hard negative sampling approach poses a significant challenge within the realm of self-supervised contrastive learning due to the absence of alternative means to describe semantics beyond model representations. However, the advent of multimodal contrastive learning offers a promising solution. Multimodal contrastive learning aims to align views from different modalities of the same instance within a unified representation space, ensuring that each modality reflects the intended semantics in its unique way. When a modality possesses a well-defined similarity measure, it enables the bypassing of model representations for hard negative sample selection, effectively serving as a \textbf{proxy for semantics.}.

Leveraging the manifestation modality as such a proxy, we have explored the distribution of semantic differences via the Hamming distance within a uniformly sampled minibatch. Crucially, since manifestations remain invariant throughout model optimization, we can \emph{extend the Hamming distance calculation across the entire dataset, thereby circumventing the minibatch limitation. This extension allows for the direct sampling of hard negatives based on a specific distribution, embodying the essence of the ManiNeg approach.}

ManiNeg, through its innovative use of an additional manifestation modality as a semantic proxy, adeptly overcomes the challenges associated with hard negative sampling based on representations. This strategy ensures a high degree of semantic alignment at the data level, mitigating the risk of selecting poor hard negative samples due to misaligned representations. Most importantly, while the proportion of hard negative samples in a uniformly sampled minibatch might be minimal, ManiNeg enables the precise and targeted identification of hard negative samples across the entire dataset, thereby addressing the issue of sampling efficiency at its core.

\subsection{ManiNeg in Implementation}
\label{sec:method:maniNegImple}

\textbf{Sampling Scheme.} 1) Initial Sampling. An anchor instance is randomly selected from the dataset without replacement, ensuring each instance has an equal chance of chosen. 2) Hamming Distance Calculation. For the chosen anchor, the Hamming distances to all other instances in the dataset are computed, providing a measure of similarity. 3) Distribution Definition. A truncated Gaussian distribution is defined over the Hamming distances. From this distribution, several distances are sampled, reflecting a controlled variability in the degree of dissimilarity from the anchor. 4) Instance Selection. For each sampled Hamming distance, an instance is uniformly selected from the group of instances that match the distance. This step ensures diversity in the sampled instances, which are then combined to form a minibatch for training.

The probability density function of the truncated Gaussian distribution defined in Step 3) is compued by:
\begin{equation}
    f(x; \mu, \sigma^2, a,b) = \begin{cases} 
    \frac{\phi(x; \mu,\sigma^2)}{\Phi(b; \mu, \sigma^2) - \Phi(a; \mu, \sigma^2)} , & \text{if } a \leq x \leq b \\
    0, & \text{others}
    \end{cases} 
    \label{eq:truncated}
\end{equation}
where $x$ represents the the Hamming distance, $\Phi$ and $\phi$ are the cumulative distribution function and the probability density function of the Gaussian distribution $\mathcal{N}(\mu, \sigma^2)$. $a$ and $b$ are the upper and lower bounds of the truncated Gaussian distribution.

The lower bound $a$, set at 1, ensures exclusion of the anchor instance from being re-sampled as a negative sample. The upper bound $b$, determined by analyzing the maximum Hamming distance across the entire dataset, is set at 18. This range encapsulates the spectrum of semantic dissimilarities within the dataset. The mean $\mu$ and standard deviation $\sigma$ act as adjustable hyperparameters that influence the variance in sample difficulty. These settings allow for a dynamic adjustment based on direct observations from sampling outcomes, simplifying the hyperparameter tuning. The distribution, referred to as the \emph{ManiNeg} curve, is visually represented in Fig.~\ref{fig:main}, serving as a graphical reference to the underlying principles of this sampling methodology.

\textbf{Manifestation Deduplication.} 
To prevent redundancy in the minibatch, instances with identical manifestations are filtered to retain only one, underlining the principle that samples with no semantic differences are not suitable as negative samples in contrastive learning.

% 注意统一用语：负样本的“难度”应该用词“hardness”
\textbf{Hardness Annealing.} 
The training regimen incorporates a hardness annealing strategy, gradually increasing the difficulty of the negative samples as the model becomes more adept. This approach allows the model to initially grasp broader features before honing in on more nuanced distinctions, with the flexibility to adjust $\mu$ in \eqref{eq:truncated} within the truncated Gaussian distribution facilitating this progression.

\textbf{Pseudocode.} 
To summarize, the complete training process is summarized in Algorithm ~\ref{alg:training}.

The computationally intensive steps, particularly those involving pre-calculable distances, can be pre-processed and stored, streamlining the training by eliminating the need for real-time computation of distances for every training step. This pre-computation significantly enhances the practicality and efficiency of ManiNeg, making it a viable and effective method for leveraging hard negative samples.
% 在这里，我们讲围绕ManiNeg的核心思路来讲述算法上的细节 

% % 可选，看能不能证出来
% Update 240102应该不重要。直接删了即可
% \subsection{Distribution of Batch}
% \label{sec:method:maniNegBatchDist}

\subsection{Further Demonstrations on ManiNeg}
\label{sec:method:demo}
The ManiNeg sampling approach, predicated on Hamming distances, offers a visually interpretable method to demonstrate its sampling and annealing process. This visualization can be understood from two distinct angles.

\begin{figure*}[t]
    \centering
    (a)
    \includegraphics[width=0.97\linewidth]{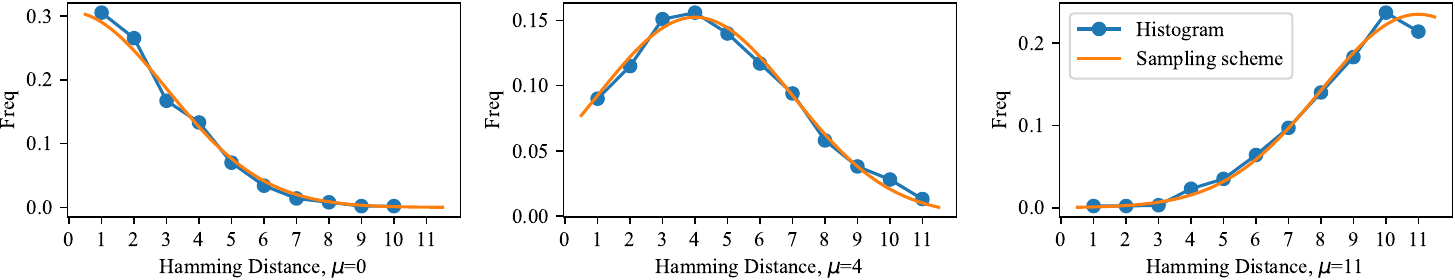} 
    % \hspace{1in}
     \\
    (b)
    \includegraphics[width=0.97\linewidth]{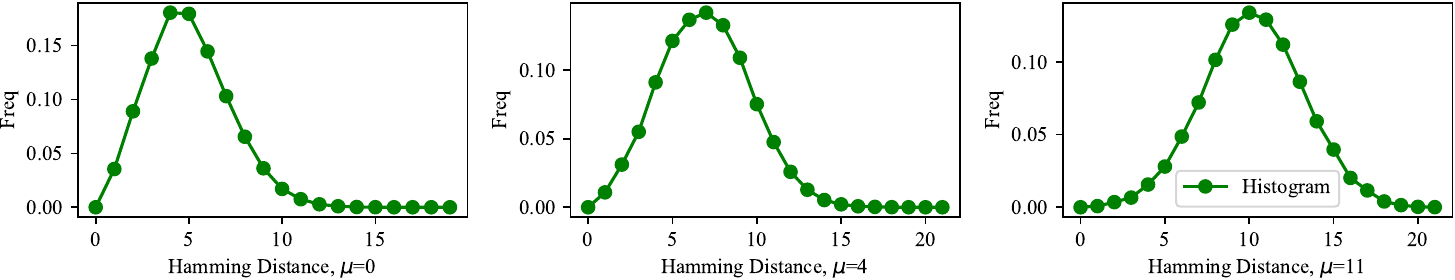}
    \caption{Demonstration of ManiNeg using different values of $\mu$ for sampling. (a) Histogram of the Hamming distances between the anchor and the negative samples. (b) Histogram of the Hamming distances between all negative pairs within the minibatch.}
\label{fig:demo}
\end{figure*}

\textbf{Distance between the anchor and the negatives.} 
The Hamming distances between the anchor and negative samples directly correspond to the outcomes sampled according to the specified truncated Gaussian distribution. Figure~\ref{fig:demo}(a) showcases minibatches sampled under various hardness settings ($\mu$), illustrating the annealing process and the distribution of the sampling scheme for reference.

\definecolor{ForestGreen}{RGB}{118,171,95}
\begin{algorithm}[H]
\caption{Python-style Pseudocode of ManiNeg.}\label{alg:training}
\begin{algorithmic}
\STATE \textcolor{ForestGreen}{\# Variable initialization}
\STATE dataset=mvkl.dataset(batchsize=1).shuffle()
\STATE batchsize
\STATE t\_gaussian \textcolor{ForestGreen}{\# Eq.~\eqref{eq:truncated}}
\STATE model
\STATE \textcolor{ForestGreen}{\# Training process}
\STATE \textbf{for} anchor \textbf{in} dataset:
\STATE \hspace{0.5cm} sample\_space = defaultdict(list)
\STATE \hspace{0.5cm} \textbf{for} neg \textbf{in} dataset: \textcolor{ForestGreen}{\# pre-computable}
\STATE \hspace{0.5cm} \hspace{0.5cm} \textbf{if} neg == anchor:
\STATE \hspace{0.5cm} \hspace{0.5cm} \hspace{0.5cm} \textbf{continue}
\STATE \hspace{0.5cm} \hspace{0.5cm} h\_d = Hamming\_distance(anchor, neg)
\STATE \hspace{0.5cm} \hspace{0.5cm} sample\_space[h\_d].append(neg)
\STATE \hspace{0.5cm} t\_gaussian.annealing() \textcolor{ForestGreen}{\# according to the training step}
\STATE \hspace{0.5cm} h\_d\_samples = t\_gaussian.sample(size=batchsize-1)
\STATE \hspace{0.5cm} batch = list()
\STATE \hspace{0.5cm} batch.append(anchor)
\STATE \hspace{0.5cm} \textbf{for} h\_d \textbf{in} h\_d\_samples:
\STATE \hspace{0.5cm} \hspace{0.5cm} sample = random.sample(sampling\_space[h\_d],1)
\STATE \hspace{0.5cm} \hspace{0.5cm} batch.append(sample)
\STATE \hspace{0.5cm} manifestation\_deduplicate(batch)
\STATE \hspace{0.5cm} model.train(batch)
\end{algorithmic}
\end{algorithm}

% The distribution of ManiNeg, the results of the sampling as well as the annealing process are depicted in Fig.~\ref{fig:demo}(a) for reference.

% 注意统一用语，重要：在不需要Loss作用在view上还是作用在Instance上时，我们并不需要区分sample和instance。但是这可能会导致行文中出现sample和instance混用。需要注意这一点

\textbf{Distance between all negative pairs.}
Efficiency precludes the exclusive selection of the anchor as the positive sample within a minibatch. Adhering to standard contrastive learning, each sample within the batch is treated as a positive instance, with the loss function \eqref{eq:clbase} computed individually and the minibatch loss determined by averaging these individual losses. Thus, the focus shifts to the distribution of Hamming distances among all negative pairs within the batch.

A formal analysis of this distribution is challenging due to its derivation from multiple overlapping distributions, which are significantly influenced by the dataset's characteristics. For instance, traits within real manifestations often exhibit dependencies, and the distribution of each trait can vary widely, complicating the derivation of a generalized formal expression. Despite these challenges, the distribution can be empirically visualized through sampling experiments, as depicted in Fig.~\ref{fig:demo}(b). This visualization reveals that increasing the hardness of negative samples (i.e., decreasing $\mu$ in \eqref{eq:truncated}) correlates with a reduction in the average Hamming distance between negative pairs, which is consistent with expectations.

This analysis underscores two critical insights. First, it highlights a pragmatic approach for selecting hyperparameters in the truncated distribution. By evaluating the effects of different hyperparameters through sampling and visualizing in Fig.\ref{fig:demo}(b), we can fine-tune the setup without engaging in model optimization, streamlining the hyperparameter selection process. Second, this methodology serves as a preliminary assessment tool for determining a dataset's compatibility with ManiNeg. The success of ManiNeg is intricately linked to the dataset's distribution, which can be complex and dataset-specific. Thus, a decreasing trend in the mean Hamming distance among negative pairs, as observed in Fig.\ref{fig:demo}(b), suggests the dataset's potential suitability for ManiNeg.

% 注意统一用语：跨数据集对应cross-dataset。注意下前文中是否有其他表述，有的话需要统一。
\section{Experiments and Results}
\label{sec:expr}
We first introduce the experimental setup and evaluation protocols, and then validate ManiNeg in unimodal, multimodal, and cross-dataset scenarios.
\subsection{Experimental Setups}
\label{sec:expr:setup}

% 注意这一流程中表征的说法（如：learns to generate representation of the input data）
% 这句话还有一种说法，即提取特征（extract feature）
% 这些说法应该与前文介绍对比学习的地方一致

% 注意统一用语：下游任务的名称，良恶性预测。这一说法应当统一。目前倾向于统一为benign-malignant **prediction**。
% 这一段应该与IV-A中的Basic framework的介绍一致
% 注意统一用语：预训练“阶段”、下游任务“阶段”中的“阶段”需要统一。有stage和phase两种说法，倾向于使用phase。
\textbf{Experimental Procedure.}
Our experimental framework is detailed in the \emph{Basic Framework} section outlined in Section \ref{sec:method:clRecap}. During pretraining, models are developed using contrastive learning in both unimodal and multimodal settings, with the downstream task focused on the benign-malignant classification of breast lumps. Reflecting practical application requirements, only the image modality is employed in the downstream task. Given the small and concealed nature of breast lumps, the capability of models to extract relevant features is a critical factor in the evaluation.

% 统一用语：单模态的表征提取，倾向于使用representation generation。这一用语也应该被统一
% 统一公式格式：注意向量要加\bm
\textbf{Modalities.}
The application of contrastive learning spans unimodal representation generation and multimodal representation alignment. While ManiNeg introduces an additional manifestation modality as a semantic proxy—impacting only the instance sampling procedure—the core pretraining of models remains applicable under both unimodal and multimodal conditions. To ensure a thorough assessment of ManiNeg, we examine models pretrained under these varied scenarios.

In the unimodal context, instances comprise solely of different mammogram views $\bm x^{cc}_i$ and $\bm x^{mlo}_j$, with $i$ and $j$ denote their origin from the same instance. The model structure is depicted in Fig.~\ref{fig:model}(a), with the substitution of $\bm z^{cc}_i$, $\bm z^{mlo}j$ into \eqref{eq:clbase} resulting in the unimodal loss function $\ell{uni}$ expressed as
\begin{equation}
    \label{eq:unimodal}
    \ell_{uni} = \ell(\bm z^{cc}_i,\bm z^{mlo}_j)~.
\end{equation}

In multimodal settings, \citet{hager2023best} has demonstrated the feasibility of integrating tabular data, such as manifestations, as an additional modality. Manifestations $\bm x^M_j$ are mapped into a shared representation space $\bm z^M_j$ utilizing a shallow neural network $f_M(\cdot)$ and a shared projector $g(\cdot)$, facilitating closer alignment with mammogram representations $\bm z^I_i$. This approach maintains image unimodal contrastive learning to enhance image feature extraction, with the network's configuration illustrated in Fig.~\ref{fig:model}, and the multimodal loss function $\ell_{multi}$ defined as
\begin{equation}
\begin{split}
    \label{eq:multimodal}
    &\ell_{M} = \frac{1}{2}(\ell_{inter}(\bm z^{cc}_i,\bm z^{M}_j)+\ell_{inter}(\bm z^{mlo}_i,\bm z^{M}_j))~, \\
    &\ell_{multi} = \ell_{M}+\ell_{uni}~.
\end{split}
\end{equation}

Here, $\ell_{inter}$ mirrors the form of \eqref{eq:clbase} but excludes negative sample pairs from identical modalities, thereby omitting pairs such as $\{z_i^{cc}, z_k^{cc}|k \neq i\}$ from the denominator of \eqref{eq:clbase}. We use ResNet50 \cite{resnet} as the feature extractor $f_I(\cdot)$, while manifestations are processed through two linear layers. The shared projector $g(\cdot)$ consists of two linear layers, fostering a cohesive and efficient learning environment across modalities.

\textbf{Datasets.}
The MVKL dataset, detailed in Section \ref{sec:dataset}, was partitioned into training, validation, and test subsets following a 7:1:2 ratio, ensuring that mammograms from the same patient remained within the same subset. The specifics of this distribution are cataloged in Table~\ref{tab:label_dist} for reference.

Subsequent validations of the downstream task were conducted on two datasets. Initially, the MVKL dataset, identical to the pretraining phase, was utilized, employing pathology-verified labels as the ground truth. The dataset's division scheme was maintained consistently throughout both the pretraining and validation phases.

To evaluate the generalization capacity, validations were extended to the CBIS-DDSM dataset \cite{lee2017curated}, which comprises 3,568 mammogram cases marked with breast lumps categorized as either mass or calcification. Each lump is accompanied by a pathology-verified label, mirroring the ground truth approach adopted for the MVKL dataset. The CBIS-DDSM dataset was divided according to its original scheme, with an additional 10\% of the training set allocated for validation purposes. The division details are also presented in Table~\ref{tab:label_dist}.

\textbf{Metrics.}
To mitigate the potential biases induced by label imbalance or threshold adjustments, area under the receiver operating characteristic (AUC)  was used for evaluation.

% 注意所有的section都要大写
\textbf{Baseline Methods.}
Two benchmark methods were selected for comparative analysis against ManiNeg. The first, referred to as \emph{uniform}, follows the original SimCLR approach detailed in the \emph{hard negative sampling scheme} part of Section \ref{sec:method:clRecap}. This method is emblematic of numerous contrastive learning studies that do not explicitly tackle the challenge of negative sample hardness. The second comparative method, stemming from the research of \citet{robinson2020hard}, employs a von Mises-Fisher distribution centered around the anchor sample for hard negative sampling, thereby favoring the selection of negative samples closer to the positive sample. This approach is denoted as \emph{vMF} in subsequent discussions.

\textbf{Data Augmentation.}
To augment the mammography images, a series of transformations were applied: \emph{RandomResizedCrop} within the scale of (0.5, 1), \emph{RandomHorizontalFlip} with a 50\% probability, and two iterations of \emph{RandAugment} \cite{cubuk2020randaugment}, excluding any color-related augmentations. These augmentations were consistently applied across all mammogram-related experiments. For manifestations, no data-level augmentations were introduced; instead, a \emph{Dropout} with a 50\% probability was applied to the manifestation features $\bm y_M$.

\textbf{Hyperparameters.}
Detailed hyperparameter settings are documented in Table~\ref{tab:hyper}, providing a transparent overview of the experimental configurations.

\renewcommand{\arraystretch}{1.3}
\begin{table}[ht]
    \centering
    \setlength{\tabcolsep}{7pt}
    \caption{Hyperparameters. }
    \begin{threeparttable}
    \begin{tabular}{l c}
        \hline
        \hline
        \textbf{Hyperparameters} & \textbf{Value} \\
        \hline
        \multicolumn{2}{c}{Pretraining}\\
        \hline
        Peak learning rate & 1e-4 \\
        Minimal learning rate & 1e-7 \\
        Learning rate schedule & Cosine annealing \\
        Training steps &  9000 \\
        Warmup steps &  300 \\
        Batch size &  64 \\
        % Adam $\beta$ &  (0.9, 0.999) \\
        Temperature $\tau$  &  trainable, initialized by 0.7 \\
        Weight decay & 1e-4\\
        Input resolution & $256^{2}$\\
        ManiNeg, maximum $\mu$& 11\\
        ManiNeg, minimum $\mu$& 0\\
        ManiNeg, $\sigma$&3\\
        ManiNeg, annealing schedule&Linear\\
        ManiNeg, annealing, step at minimum $\mu$\tnote{$\dagger$}&150th\\

        \hline
        \multicolumn{2}{c}{Downstream, linear evaluation}\\
        \hline
        Peak learning rate &  1e-3 \\
        Minimal learning rate &  1e-6 \\
        Learning rate schedule &  Cosine annealing \\
        Maximum epochs &  1000 \\
        Early stopping patience & 100 \\
        Batch size &  48 \\
        % Adam $\beta$ &  (0.9, 0.999) \\
        Weight decay & 1e-6\\
        Input resolution & $256^{2}$\\
        \hline
        \multicolumn{2}{c}{Downstream, fine-tuning}\\
        \hline
        Peak learning rate  & 5e-5 \\
        Minimal learning rate &  5e-7 \\
        Learning rate schedule &  Cosine annealing \\
        Layer-wise learning rate decay\tnote{$\ddagger$}  &  0.1 \\
        Maximum epochs &  1000 \\
        Early stopping patience & 100 \\
        Batch size &  48 \\
        % Adam $\beta$ &  (0.9, 0.999) \\
        Weight decay & 5e-5\\
        Input resolution & $256^{2}$\\
        \hline
        \multicolumn{2}{c}{Downstream, linear probe}\\
        \hline
        $\mathcal{L}_2$ regularization strength $\lambda$  & 3.16 \\
        Maximum iteration &  1000 \\
        \hline
        \hline
    \end{tabular}
    
     \begin{tablenotes}
        \footnotesize
        \item[$\dagger$] During the training process, $\mu$ starts from a maximum value of 11 and stops at a minimum value of 0 at the 150th training step, remaining unchanged thereafter. The decrease in $\mu$ is linear.
        \item[$\ddagger$] The learning rate on the pretrained parameters is 0.1 times that of the learning rate on the appended randomly initialized linear layer.
      \end{tablenotes}
    \end{threeparttable}
\label{tab:hyper}
\end{table}

% 统一用语：fine-tune有连字符
% 绘图时绘制下游任务的图
\subsection{Evaluation Protocol}
\label{sec:expr:protocol}
Two primary methodologies are considered to evaluate the downstream tasks. The first method involves freezing the pretrained model's parameters to directly evaluate the generated representations. The second method employs the pretrained model as a starting point, upon which further fine-tuning is conducted for the downstream task.

In alignment with established methodologies from SimCLR~\cite{chen2020simple} and CLIP~\cite{radford2021learning}, three distinct evaluation protocols are utilized for downstream task assessment. Given the exclusive use of the image modality for downstream task validation, the pretrained models referenced herein specifically pertain to the pretrained image branch $f_I(\cdot)$.

\textbf{Linear Probe (LP)} maintains the pretrained model's parameters and employs logistic regression to categorize the representations into benign or malignant classes.

\textbf{Linear Evaluation (LE)} preserves the pretrained model's parameters but introduces an additional trainable linear layer for classifying the representations.

\textbf{Fine-tuning (FT)} preserves the pretrained model's parameters in a fixed state but introduces an additional trainable linear layer for classifying the representations.

% 注意：模型以ImageNet作为初值，而不是From Scratch作为初值
\subsection{Evaluation Results}
\label{sec:expr:result}
We evaluate the methods on MVKL and CBIS-DDSM under both unimodal and multimodal pretraining scenarios. All the experiments are repeated 10 times using the same hyperparameters and different random seeds. The mean and the standard deviation are calculated base on these results.

\textbf{Unimodal.} The results on MVKL are shown in Table~\ref{tab:result:uni_mvkl}, and the results on CBIS-DDSM are recorded in Table~\ref{tab:result:uni_ddsm}. The challenge in extracting representations of breast lumps, owing to their small and concealed nature, underscores the limitations of self-supervised learning in isolating critical lesion characteristics without lesion-specific information. Despite these challenges, ManiNeg demonstrates a performance that is slightly superior or comparable to the best outcomes among the baseline methods. This suggests that ManiNeg's strategy of explicitly sampling hard negative samples endows it with enhanced robustness against the baseline approaches.

\renewcommand{\arraystretch}{1.5}
\begin{table}[ht]
    \centering
    \setlength{\tabcolsep}{8pt}
    \caption{Unimodal Pretraining Results on MVKL (mean$\pm$std).}
    \begin{tabular}{l c c c}
        \hline\hline
         & LP & LE & FT  \\
        \hline
        Uniform & 57.71$\pm$1.75 & 72.47$\pm$0.64 & 76.38$\pm$0.72 \\
        vMF \cite{robinson2020hard} & 60.45$\pm$1.23 & 71.72$\pm$0.75 & 77.53$\pm$0.19 \\
        \hline
        ManiNeg & \textbf{60.46$\pm$1.22} & \textbf{72.73$\pm$0.27} & \textbf{77.54$\pm$1.15}  \\
        \hline\hline
    \end{tabular}
    \label{tab:result:uni_mvkl}
\end{table}

\renewcommand{\arraystretch}{1.5}
\begin{table}[ht]
    \centering
    \setlength{\tabcolsep}{8pt}
    \caption{Unimodal Pretraining Results on CBIS-DDSM (mean$\pm$std).}
    \begin{tabular}{l c c c}
        \hline\hline
         & LP & LE & FT  \\
        \hline
        Uniform & 63.38$\pm$0.70 & 67.06$\pm$0.45 & 72.33$\pm$1.02  \\
        vMF \cite{robinson2020hard} & 60.58$\pm$1.06 & 65.46$\pm$0.52 & 72.00$\pm$0.45 \\
        \hline
        ManiNeg & \textbf{64.64$\pm$0.91} & \textbf{67.44$\pm$1.56} & \textbf{72.92$\pm$0.82}  \\
        \hline\hline
    \end{tabular}
    \label{tab:result:uni_ddsm}
\end{table}

\textbf{Multimodal.} Given that ManiNeg integrates manifestation data during the batch sampling phase, it logically extends to include this modality in the pretraining process as well. The outcomes of multimodal pretraining are available in Tables~\ref{tab:result:multi_mvkl} and \ref{tab:result:multi_ddsm}, representing evaluations on the MVKL and CBIS-DDSM datasets, respectively.

Incorporating manifestation data into pretraining enables the model to identify lesion-specific information within the images, leveraging the manifestations as a guide. The results indicate a significant performance uplift with ManiNeg in this multimodal setting compared to baseline methods, highlighting the added value of multimodal integration.

An additional insight from these evaluations is the performance of the vMF method relative to the uniform sampling approach. Although the vMF was conceived as an enhancement over the uniform method, it occasionally produces inferior outcomes across several tests. The vMF approach, despite its intent to refine the selection of negative samples within a uniformly sampled minibatch, can, based on its hyperparameter settings, apply a more aggressive weighting to negative samples than what is seen with cross-entropy weighting in uniform sampling. This aggressive approach could exacerbate the challenges related to misalignment between representation and semantics, as well as sampling efficiency. The empirical data from these experiments supports this hypothesis, illustrating the nuanced impact of different negative sampling strategies on model performance.

\renewcommand{\arraystretch}{1.5}
\begin{table}[ht]
    \centering
    \setlength{\tabcolsep}{8pt}
    \caption{Multimodal Pretraining Results on MVKL (mean$\pm$std).}
    \begin{tabular}{l c c c}
        \hline\hline
         & LP & LE & FT  \\
        \hline
        Uniform & 63.52$\pm$0.82 & 73.78$\pm$0.67 & 77.75$\pm$0.37  \\
        vMF \cite{robinson2020hard} & 59.11$\pm$1.16 & 72.01$\pm$0.69 & 76.44$\pm$0.41 \\
        \hline
        ManiNeg & \textbf{65.16$\pm$1.26} & \textbf{74.36$\pm$0.35} & \textbf{79.82$\pm$0.51}  \\
        \hline\hline
    \end{tabular}
    \label{tab:result:multi_mvkl}
\end{table}

\renewcommand{\arraystretch}{1.5}
\begin{table}[ht]
    \centering
    \setlength{\tabcolsep}{8pt}
    \caption{Multimodal Pretraining Results on CBIS-DDSM (mean$\pm$std).}
    \begin{tabular}{l c c c}
        \hline\hline
         & LP & LE & FT  \\
        \hline
        Uniform & 60.88$\pm$0.88 & 67.47$\pm$0.70 & 70.51$\pm$0.71  \\
        vMF \cite{robinson2020hard} & 63.78$\pm$1.57 & 64.03$\pm$1.40 & 70.60$\pm$0.55 \\
        \hline
        ManiNeg & \textbf{64.71$\pm$0.93} & \textbf{70.54$\pm$0.90} & \textbf{74.55$\pm$0.99}  \\
        \hline\hline
    \end{tabular}
    \label{tab:result:multi_ddsm}
\end{table}

\begin{figure*}[ht]
    \centering
    \includegraphics[width=0.97\linewidth]{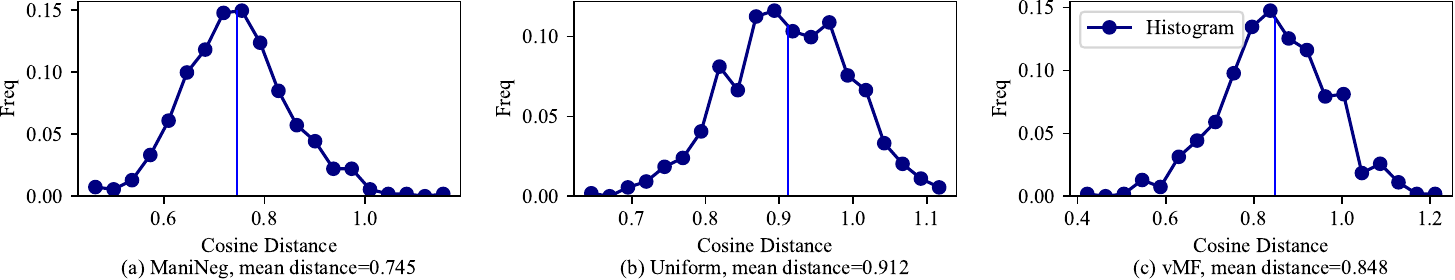} 
    \caption{Multimodal alignment illustrations. The models pretrained in the multimodal scenario with (a) ManiNeg, (b) uniform, and (c) vMF are used to plot histograms of the cosine distances between image representations and manifestation representations for all positive pairs in the test set. The mean distance is noted in the histogram titles and marked on the histograms with blue vertical lines.}
\label{fig:align}
\end{figure*}

\textbf{Multimodal Alignment.} Ideal hard negative samples play a pivotal role in enhancing a model's feature extraction capabilities, leading to more informative representations. This benefit is particularly evident through modality alignment, where models pretrained with ManiNeg, uniform, and vMF approaches are assessed for their ability to align image and manifestation representations. In experiments, the cosine distance between image and manifestation representations of all positive pairs in the test set was measured, with results plotted in Fig.~\ref{fig:align}. These findings indicate that ManiNeg achieves the smallest inter-modal distance, suggesting that the image branch of the model has effectively assimilated more information from the manifestations, thereby improving the precision in localizing and analyzing lesions.

\section{Discussions}
\label{sec:discussion}
While ManiNeg demonstrates theoretical and empirical advantages, a consideration of its application scenarios is crucial for its broader adoption across datasets and tasks. ManiNeg, similar to the uniform method, predicates on certain assumptions regarding the data, especially concerning the semantic proxy, such as manifestations. For the proxy's differences to meaningfully reflect semantic distinctions, the semantic proxy must be thoughtfully curated. Furthermore, given ManiNeg's reliance on Hamming distance—which treats disparities in the semantic proxy uniformly—each aspect of the proxy should contribute equitably to the downstream task's objectives. A scenario where a specific trait directly correlates with the task's outcome (often the 'gold standard' in medical contexts) might render the semantic proxy's Hamming distance ineffectual, making supervised contrastive learning~\cite{khosla2020supervised} a more fitting approach. However, the trend in contrastive learning towards generating universal representations that are not narrowly tailored to a single task implies that ManiNeg's uniform treatment of semantic proxy elements could offer enhanced generalization for diverse downstream tasks. The effectiveness of ManiNeg's sampling also depends on the distribution of the semantic proxy, which is influenced by the data's intrinsic properties and the proxy's design. To ensure sampling effectiveness, the procedure illustrated in Fig.~\ref{fig:demo}(b) should be employed to validate ManiNeg's applicability to new datasets, with larger semantic proxies likely yielding benefits due to the central limit theorem.

A key goal of deep learning, and self-supervised learning in particular, is to minimize annotation costs. Although ManiNeg necessitates the additional step of annotating manifestations—a task not typically included in clinical workflows—the medical training required for accurate annotation ensures that this process remains feasible, albeit potentially limiting ManiNeg's large-scale application. Thus, for extensive pretraining across a broad spectrum of tasks with ample data, utilizing large batch sizes and the uniform method may be practical. Conversely, in specialized fields with constrained data volumes, where the inefficiency of hard negative sampling becomes a significant issue, the annotation costs become justifiable, positioning ManiNeg as a preferable solution.

Future developments for ManiNeg will focus on extending its application to diverse datasets and tasks, including those on a larger scale. Addressing the challenges of designing and acquiring a suitable semantic proxy modality will be critical. Data-driven approaches may offer solutions to these challenges, warranting further exploration into ManiNeg's robustness against noise and its adaptability beyond tabular data to other accessible modalities and similarity measures. These considerations will guide the direction of ManiNeg's ongoing evolution and its potential broadening to encompass a wider array of applications.

\section{Conclusion}
\label{sec:conclusion}
The development of ManiNeg is aimed at addressing specific challenges encountered in the self-supervised contrastive learning process for mammographic imaging. In this study, we introduce ManiNeg, a novel hard negative sample sampling scheme designed for mammographic contrastive learning. Diverging from traditional uniform sampling methods, ManiNeg leverages the manifestation modality as a semantic proxy, enabling direct sampling of hard negative samples. This innovative approach effectively tackles the potential misalignment between representations and semantics, as well as the efficiency issues inherent in uniform sampling. Demonstrated through both sampling demonstrations and downstream task validations, ManiNeg's sampling scheme has shown significant advantages, highlighting its effectiveness in enhancing model learning and performance.

We firmly believe that the concept of directly sampling hard negative samples holds substantial promise for advancing the field of contrastive learning, particularly in medical imaging contexts. Consequently, our future endeavors will concentrate on expanding the application of ManiNeg across a broader spectrum of datasets and tasks. Additionally, we will explore the utilization of different semantic proxies to further refine and adapt the ManiNeg approach. By continuing to evolve ManiNeg, we aim to unlock new potentials in self-supervised learning, paving the way for more accurate and efficient diagnostic tools in mammography and beyond.

% 这个表也是vikl里抄的。需要更新，需要视情况修改 Done

{
\renewcommand*{\bibfont}{\normalfont\footnotesize}
\printbibliography

}
\end{document}